\documentclass[prb, twocolumn, 
showkeys,amsmath,amsfonts,floatfix,superscriptaddress,nofootinbib,longbibliography, 10pt, aps]{revtex4-2}
\usepackage[normalem]{ulem}
 
\usepackage{amssymb,amsmath,amstext}               
\usepackage{graphicx}   
\usepackage{lineno}                                    
\usepackage{bm}           
\usepackage{appendix}       
\usepackage{latexsym}
\usepackage[version=4]{mhchem}
\usepackage[colorlinks=true,citecolor=blue,linkcolor=magenta]{hyperref}
\usepackage{nameref} 
\usepackage{xcolor}
\usepackage{microtype}

\usepackage{changes}

\def\be{\begin{equation}}
\def\ee{\end{equation}}
\def\bea{\begin{eqnarray}}
\def\eea{\end{eqnarray}}
\def\bi{\begin{itemize}}
\def\ei{\end{itemize}}

\newcommand{\bra}[1]{\mbox{$\langle #1 |$}}
\newcommand{\ket}[1]{\mbox{$| #1 \rangle$}}

\makeatletter
\def\moverlay{\mathpalette\mov@rlay}
\def\mov@rlay#1#2{\leavevmode\vtop{%
   \baselineskip\z@skip \lineskiplimit-\maxdimen
   \ialign{\hfil$\m@th#1##$\hfil\cr#2\crcr}}}
\newcommand{\charfusion}[3][\mathord]{
    #1{\ifx#1\mathop\vphantom{#2}\fi
        \mathpalette\mov@rlay{#2\cr#3}
      }
    \ifx#1\mathop\expandafter\displaylimits\fi}
\makeatother

\definecolor{norange}{RGB}{230,120,20}

\begin{document}

\newcommand{\aw}[1]{{\color{red}[AW: #1]}}


\title{Evolution of superconductivity from charge clusters to stripes in the $t$--$t'$--$J$ model}

\author{Aritra Sinha}
\email{asinha@pks.mpg.de}
\affiliation{Max Planck Institute for the Physics of Complex Systems, N\"{o}thnitzer Strasse 38, Dresden 01187, Germany}

\author{Hannes Karlsson}
\affiliation{Max Planck Institute for the Physics of Complex Systems, N\"{o}thnitzer Strasse 38, Dresden 01187, Germany}

\author{Martin Ulaga}
\affiliation{Max Planck Institute for the Physics of Complex Systems, N\"{o}thnitzer Strasse 38, Dresden 01187, Germany}
             
\author{Alexander Wietek} 
\affiliation{Max Planck Institute for the Physics of Complex Systems, 
             N\"{o}thnitzer Strasse 38, Dresden 01187, Germany}

\date{\today}

\begin{abstract}
Competition and coexistence of charge orders and superconductivity are hallmarks in many strongly correlated electron systems. Here, we unravel the precise role of charge fluctuations on the superconducting state in the $t$--$t'$--$J$ model of the high-temperature cuprate superconductors. Using finite-temperature tensor network simulations, we investigate thermal snapshots in the underdoped regime where the ground state features a superconducting stripe phase. At intermediate temperatures, where stripes have melted and hole clustering is observed, we find that pairing correlations are tightly localized on the hole clusters. Upon entering the stripe regime at lower temperatures, pairing increasingly delocalizes across different hole clusters to ultimately become coherent across the full system in the ground state. This pair-charge locking gives rise to an intuitive picture of the parent state of the superconducting stripe phase: pairing is localized on hole clusters formed via hole attraction due to the onset of magnetic correlations at intermediate temperature. We discuss how this microscopic picture is consistent with a broad range of experimental observations in cuprate superconductors, including scanning tunneling microscopy (STM) evidence for local pairing above $T_c$ and nuclear magnetic resonance (NMR) signatures of charge clustering in the underdoped regime.
\end{abstract}

\maketitle
The emergence of superconductivity upon doping a Mott insulating parent state is a central topic in the study of high-temperature superconductivity. In the cuprate superconductors, the superconducting dome is accompanied by a pseudogap regime~\cite{norman2005friend}, whose electronic state of matter is intensely debated and where charge order, including stripes and charge-density modulations, across many compounds and dopings are observed~\cite{tranquada1995evidence,keimer2014high,fradkin2015,Ghiringhelli2012,chang2012direct,comin2016resonant}. 
Hence, superconductivity in the cuprates does not develop in a charge-uniform background, but in the presence of strong and often spatially inhomogeneous charge fluctuations~\cite{Pan2001,Gomes2007,Kohsaka2007,tromp2023puddle,Li2021,Bakharev2004,Vuckovic2025,Pelc2018}. Early on, it has been suggested that doped Mott systems tend toward phase separation of hole-rich and hole-poor regions, driven by competition between kinetic
energy and antiferromagnetic (AFM) exchange~\cite{emerykivelson1990,argumenthellberg}. However, for cuprates, macroscopic phase separation has been argued to be frustrated by long-range Coulomb interactions, lattice energetics, and geometry, promoting intermediate-length-scale charge order instead, most prominently stripes~\cite{emery1993frustrated}. In this picture, doped charge collects into hole-rich lines that separate AFM domains~\cite{tranquada1995evidence,zaanen1989,Kivelson2003}. An important question remains: what is the role of pairing correlations in the pseudogap regime at intermediate-temperature preceding the low-temperature stripe or superconducting orders. Does pairing reside mainly in the hole-poor AFM background, as in pairing-from-spin-fluctuations scenarios~\cite{scalapino2012commonthread,moriya2000spinfluctuations},
or are they strongest on the hole-rich regions themselves~\cite{white2009pairingstriped,hamidian2016pdw}?

In the strong-correlation regime, the Hubbard and $t$--$J$ models capture this
competition between hole motion and antiferromagnetic exchange, and numerical studies find stripes and strong pairing correlations in nearby parameter regimes~\cite{Himeda2002, wieteksquare2021,jiang2024sixleg,xu2024science,Qu2024,li2023prl,chen2021prb, Chen2025GlobalPhaseDiagram, Devereaux2025SignificanceStripes}. A key control parameter is the
next-nearest-neighbor hopping $t'$, which can substantially shift the balance between stripe order and pairing~\cite{jiang2019science,jiang2020prr,jiang2021pnas,jiang2022prb,chung2020prb,zhang2025prl,lu2023prb}.

Modern tensor-network methods allow for exploring physics at non-zero temperatures accurately, including purification-based approaches, exponential tensor renormalization group methods and cluster extensions~\cite{czarnik2012,kshetrimayum2019,sinha2022,xtrg2021quantum,zhang2025finite,DeMeyer2026Lowering}.
Here we use minimally entangled typical thermal states (METTS), which sample a Markov chain of pure states whose expectation values
approximate thermal traces and provide access to finite-temperature snapshots~\cite{white2009,stoudenmire2010minimally,wieteksquare2021,wietektriangular2021,sinha2024}.

An accurate diagnostic of pairing and condensate formation is the eigenstructure to the two-particle reduced density matrix (2RDM). As suggested by Penrose, Onsager, and Yang, the formation of a pairing condensate manifests itself by a leading eigenvalue scaling linearly with the number of particles in the system~\cite{penrose1956,yang1962,leggett2006quantum}. The eigenvectors are interpreted as the pair wave function of the Cooper pairs, which encode essential properties such as symmetry and localization of a (quasi-)condensate~\cite{Karlsson2026CooperCondensation}. While previous studies of the 2RDM mostly focused on ground state superconducting states of $t$-$J$ and Hubbard models~\cite{Wietek2022FragmentedPRL,Baldelli2025npjQM}, we here extend this analysis to finite temperatures, especially at temperatures relevant for the pseudogap regime.     

\begin{figure}[t!]
\centering
\includegraphics[width=1.0\columnwidth,clip=true]{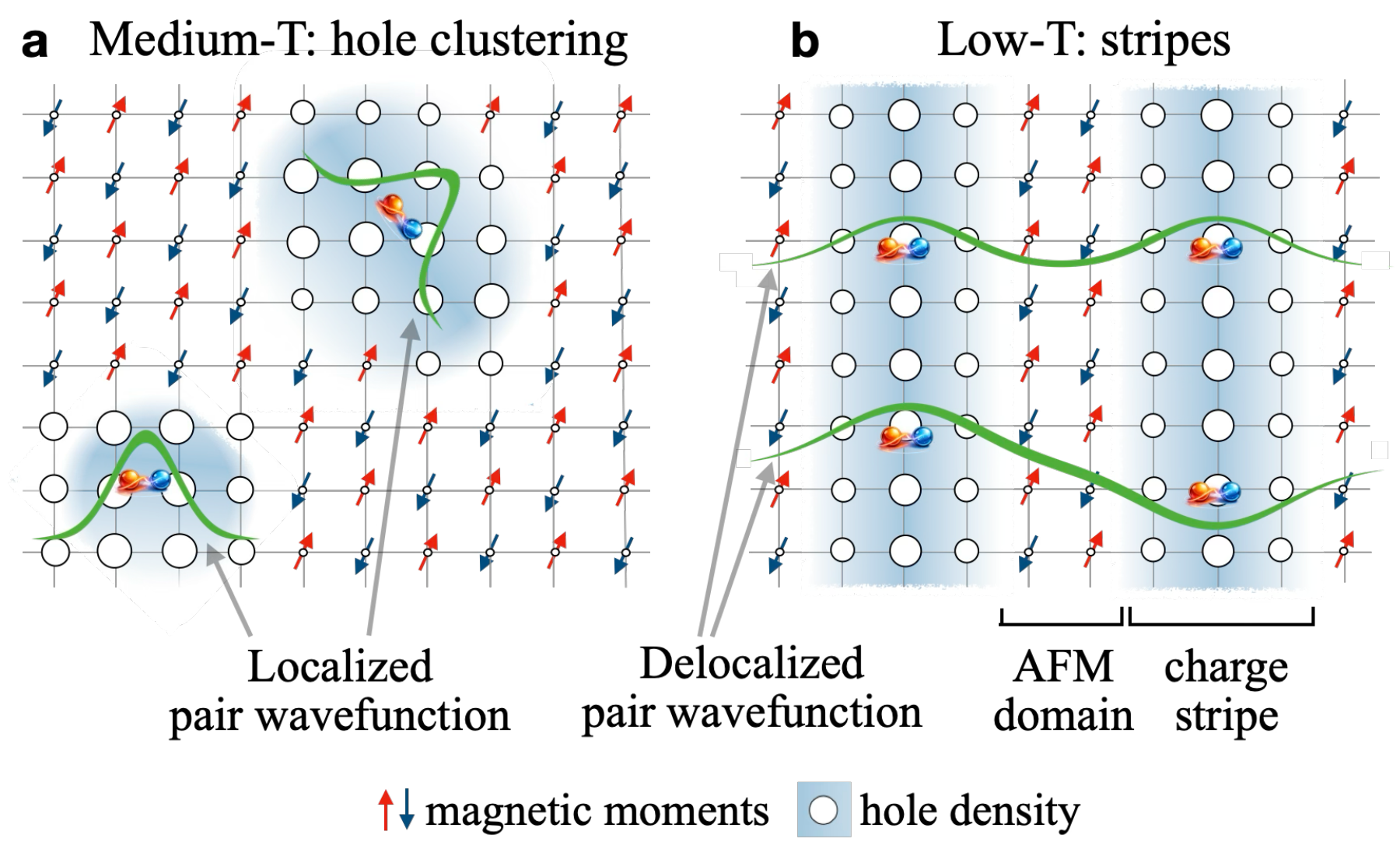}
\caption{\textbf{Schematic evolution from cluster-localized pairing to stripe-coherent superconductivity in the $t$--$t'$--$J$ model.}
\textbf{a.} At intermediate temperatures, after static stripe order has melted, doped holes form hole-rich clusters embedded in an AFM background. The dominant pair wavefunctions are localized on these hole-rich clusters. \textbf{b.} Upon further cooling, the charge clusters reorganize into stripes and the leading pair wavefunction becomes increasingly delocalized across the system, yielding a coherent $d$-wave like structure. The blue shading indicates the hole-dense regions, the red and blue arrows indicate local magnetic moments, and the green envelopes schematically indicate the spatial structure of the dominant pair wavefunctions.}
\label{fig:1}
\end{figure}

In this work, we study the $t$--$t'$--$J$ model at $J/t=0.4$, $t'/t=0.2$ and doping $p=1/16$, where previous ground-state cylinder
calculations indicate intertwined stripe and superconducting tendencies~\cite{jiang2020,Gong2021,Wietek2022FragmentedPRL}.
Using METTS, we access finite-temperature snapshots and analyze pairing and charge organization within the same microscopic states. We construct the nearest-neighbour singlet pair density matrix, resolve its leading eigenvalues and pair wavefunctions, and simultaneously
quantify charge inhomogeneity with a snapshot-level hole-cluster analysis~\cite{Sinha2025}. We find an intermediate-temperature
regime where holes form mesoscopic clusters without macroscopic phase separation. In the same temperature window, several pairing eigenvalues
are enhanced and separate from the rest of the spectrum. The corresponding pair wavefunctions are localized on the hole-rich
clusters. Upon further cooling, as stripe order emerges, the leading pair wavefunction evolves towards a coherent, extended
$d$-wave-like structures spanning the system. This physical picture is summarized schematically in Fig.~\ref{fig:1}:
pairing first nucleates on hole-rich clusters and only later acquires coherence across the stripe background.

\noindent \textbf{Model and methods.-} The $t$--$t'$--$J$ model is given by,
\begin{equation}
\begin{aligned}
H =\;&
-t \sum_{\langle i,j\rangle,\sigma}
\left(\tilde c^\dagger_{i\sigma}\tilde c_{j\sigma} + \mathrm{h.c.}\right)
-t' \sum_{\langle\!\langle i,j\rangle\!\rangle,\sigma}
\left(\tilde c^\dagger_{i\sigma}\tilde c_{j\sigma} + \mathrm{h.c.}\right)\\
&+J \sum_{\langle i,j\rangle}
\left(\mathbf S_i\cdot\mathbf S_j - \tfrac14 n_i n_j\right),
\end{aligned}
\label{eq:ttJ_ham}
\end{equation}
where $\tilde c_{i\sigma}=c_{i\sigma}\left(1-n_{i\bar\sigma}\right)$ are Gutzwiller-projected operators enforcing no double occupancy, $\langle i,j\rangle$ and $\langle\!\langle i,j\rangle\!\rangle$ denote nearest and next-nearest neighbors, $\mathbf S_i=\tfrac12\sum_{\alpha\beta} c^\dagger_{i\alpha}\,\boldsymbol{\sigma}_{\alpha\beta}\,c_{i\beta}$, and $n_i=\sum_\sigma c^\dagger_{i\sigma}c_{i\sigma}$.
We set $t=1$ and focus on $J=0.4$, $t'=0.2$ (positive $t'$, i.e., the sign convention often associated with electron-doped cuprates), and doping $p=1/16$ for the rest of the main text. Comparisons with other values of $t'$ are done in Appendix~\ref{app:dom_sector}. The system is studied on cylindrical geometries with open boundaries along $\hat{x}$ (length $L$) and periodic boundaries around $\hat{y}$ (circumference $W$). Thermal expectation values,
\begin{equation}
\langle O\rangle_T
={\mathcal Z}^{-1}\,\mathrm{Tr}\!\left(e^{-\beta H}\, O\right),
\end{equation}
$\beta=1/T$, are evaluated using the METTS algorithm~\cite{white2009, stoudenmire2010minimally}.
Starting from a product state $\ket{\sigma_s}$ we generate the snapshot states (or METTS),
\begin{equation}
\ket{\psi_s}
=\frac{e^{-\beta H/2}\ket{\sigma_s}}
{\sqrt{\bra{\sigma_s}e^{-\beta H}\ket{\sigma_s}}},
\label{eq:metts_state}
\end{equation}
measure observables in $\ket{\psi_s}$, and obtain thermal estimates from the Markov chain average
\begin{equation}
\langle O\rangle_T \approx \frac{1}{N_s}\sum_{s=1}^{N_s}\bra{\psi_s} O\ket{\psi_s}
\label{eq:metts_average}
\end{equation}
where $N_s$ denotes the number of METTS samples. Implementation details are given in Appendix~\ref{app_technical}.

\begin{figure}[t!]
\centering
\includegraphics[width=0.9\columnwidth,clip=true]{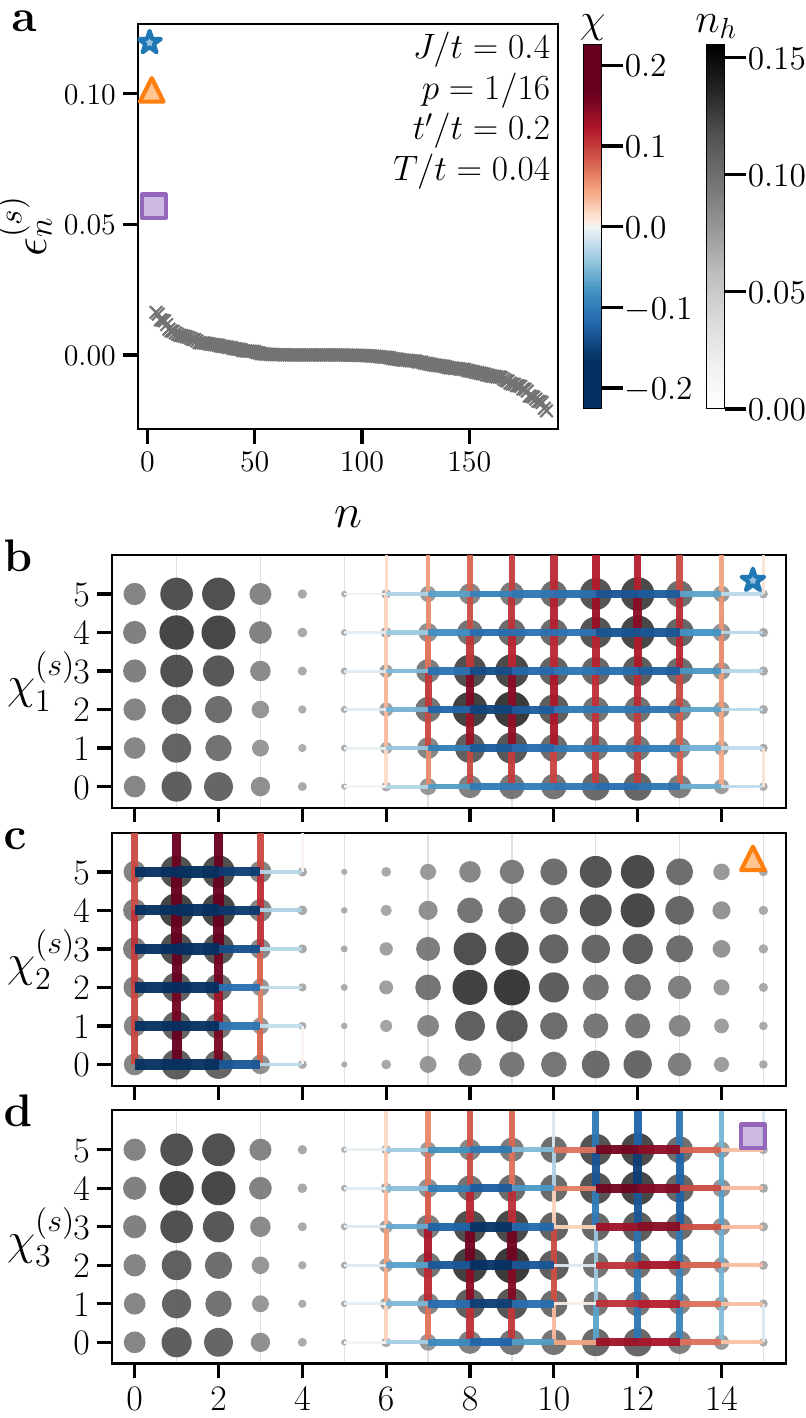}
\caption{\textbf{Pair wavefunctions in a single METTS snapshot.}
(\textbf{a}) Eigenvalues $\epsilon_n^{(s)}$ and (\textbf{b--d}) the three leading pair wavefunctions $\chi_n^{(s)}$ obtained by diagonalizing the snapshot singlet-pair correlation matrix $\rho_2^{(s)}$ [cf. Eqs.~\eqref{eq:pair_matrix_snapshot} and \eqref{eq:pair_diag}] for one representative METTS configuration for doping $p=1/16$, $J/t=0.4$ and $t'/t=0.2$ on an $L=16$, $W=6$ cylinder at $T/t=0.04$. In (\textbf{a}), the three markers identify the eigenvalues corresponding to the eigenvectors displayed in (\textbf{b--d}): the largest positive eigenvalue is marked by a blue star, the second by an orange triangle, and the third by a violet square; the subleading eigenvalues are shown as gray crosses. In (\textbf{b--d}), circles on lattice sites encode the snapshot hole density $n_h^{(s)}(\mathbf r)$ [Eq.~\eqref{hole_density_snapshot}], with both circle area and grayscale intensity increasing with $n_h^{(s)}(\mathbf r)$ (right colorbar beside \textbf{(a)}). Colored nearest-neighbour bonds encode the signed amplitude of the pair wavefunction $\chi_n^{(s)}$ (left colorbar beside \textbf{(a)}). The leading pair wavefunctions are concentrated on the hole-rich regions.}
\label{fig:2}
\end{figure}
\begin{figure*}[t]
\centering
\includegraphics[width=0.95\textwidth]{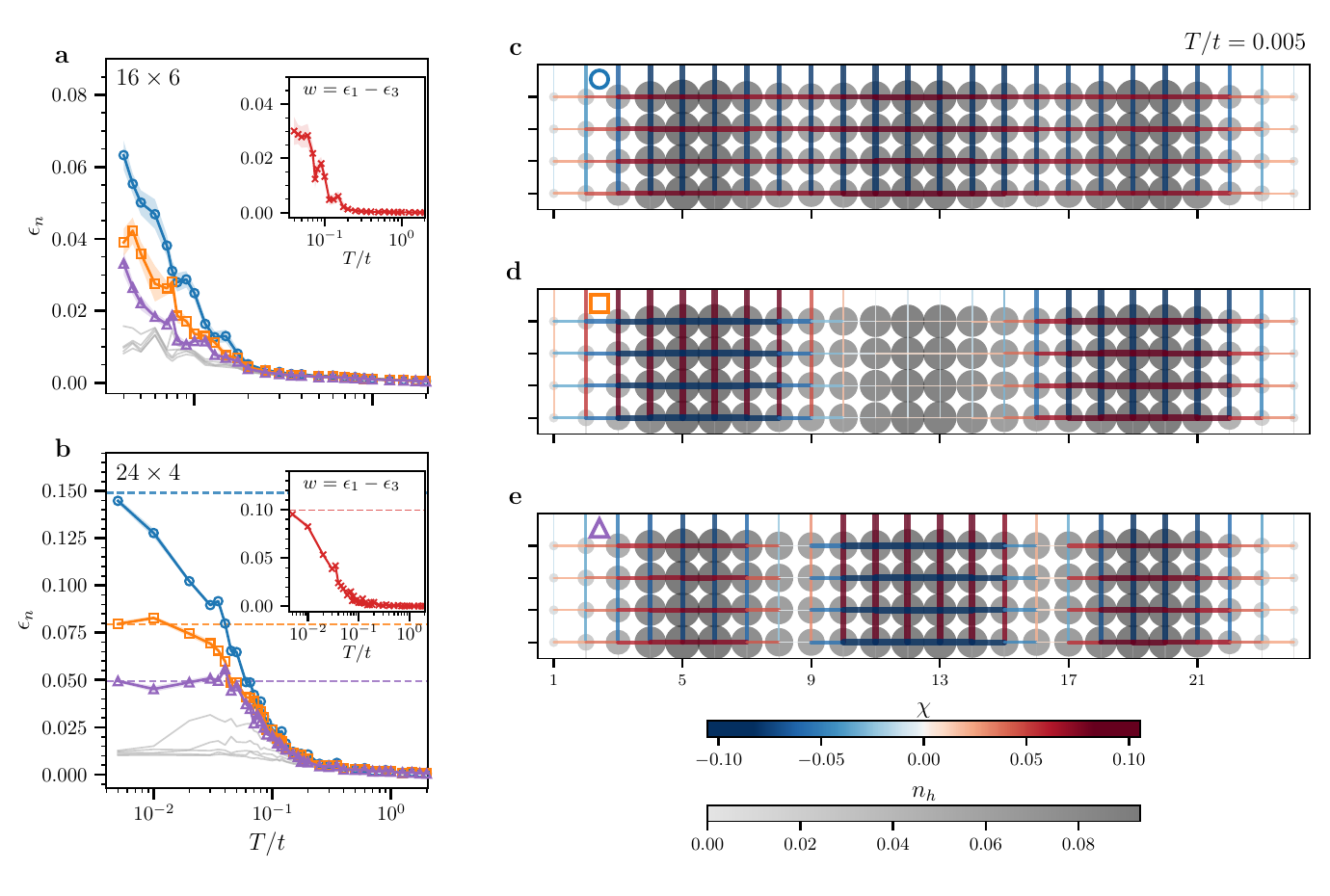}
\caption{\textbf{Thermal pairing spectrum of the $t$--$t'$--$J$ model.}
Temperature dependence of the leading eigenvalues $\epsilon_n$ of the thermally averaged singlet pairing density matrix $\rho_{2}$ (Eq.~\eqref{eq:pair_matrix}) on cylinders $16\times 6$ (panel~(\textbf{a})) and $24\times 4$ (panel~(\textbf{b})) for doping $p=1/16$, $J/t=0.4$ and $t'/t=0.2$. Curves are color-coded by eigenvalue index; the three leading eigenvalues ($n=1,2,3$) are highlighted with distinct markers, while sub-leading eigenvalues are shown in gray. The largest positive eigenvalue is marked by a blue circle, the second by an orange square, and the third by a violet triangle. Shaded bands denote statistical uncertainties from bootstrap resampling over METTS samples. Insets show the eigenvalue splitting $w=\epsilon_1-\epsilon_3$ which characterizes the Josephson coupling between the condensates, and they start increasing sharply below $T/t \lesssim0.2$. The $T/t=0$ DMRG reference is indicated by a dashed line. Right subpanels (\textbf{c--e}) display the corresponding pair wavefunctions at temperature $T/t=0.005$, together with the averaged hole-densities.}
\label{fig:3}
\end{figure*}

To diagnose pairing correlations, we analyze the eigenstructure of a
singlet-pair reduced density matrix, following the Penrose--Onsager criterion~\cite{penrose1956,yang1962,leggett2006quantum}. 
We assume nearest-neighbour singlet pairing as the dominant channel \cite{Wietek2022FragmentedPRL,jiang2021pnas}, and compute the corresponding sector of the two-particle density matrix:  for a bond
$\alpha\equiv(\mathbf r,\mu)$ with $\mu\in\{\hat{x},\hat{y}\}$ we define the singlet pairing operator,
\begin{equation}
\Delta^\dagger_{\alpha}
=\frac{1}{\sqrt{2}}
\left(
\tilde c^\dagger_{\mathbf r\uparrow}\tilde c^\dagger_{\mathbf r+\mu,\downarrow}
-
\tilde c^\dagger_{\mathbf r\downarrow}\tilde c^\dagger_{\mathbf r+\mu,\uparrow}
\right).
\label{eq:pair_operator}
\end{equation}
Our key tool for diagnosing pairing is the singlet 2RDM,
\begin{equation}
\rho_2(\alpha, \alpha')
=\big\langle \Delta^\dagger_\alpha \Delta_{\alpha'}\big\rangle_T,
\label{eq:pair_matrix}
\end{equation}
and for each snapshot, $s$, the snapshot singlet 2RDM,
\begin{equation}
    \rho_2^{(s)}(\alpha, \alpha') = \bra{\psi_s}\Delta^\dagger_\alpha \Delta_{\alpha'}\ket{\psi_s},
\label{eq:pair_matrix_snapshot}
\end{equation}
such that,
\begin{equation}
 \rho_2(\alpha, \alpha')   
\approx \frac{1}{N_s}\sum_{s=1}^{N_s}\bra{\psi_s}\Delta^\dagger_\alpha \Delta_{\alpha'}\ket{\psi_s} = \frac{1}{N_s}\sum_{s=1}^{N_s} \rho_2^{(s)}(\alpha, \alpha').
\end{equation}
To suppress strictly local density/spin contributions in this bond representation, we apply the non-overlap prescription of
Ref.~\cite{Wietek2022FragmentedPRL} and set $\rho_2(\alpha,\alpha')=\rho_2^{(s)}(\alpha,\alpha')=0$ whenever the two bonds share a common lattice site. Since $\rho_2$ and $\rho_2^{(s)}$ are Hermitian, we consider the real eigendecompositions,
\begin{align}
\rho_2(\alpha,\alpha')
&=\sum_{n}\epsilon_n\,\chi_n(\alpha)^*\,\chi_n(\alpha'),\\
\rho_2^{(s)}(\alpha,\alpha')
&=\sum_{n}\epsilon_n^{(s)}\,\chi_n^{(s)}(\alpha)^*\,\chi_n^{(s)}(\alpha').
\label{eq:pair_diag}
\end{align}
We refer to the leading eigenvalues $\epsilon_n$ (resp. $\epsilon_n^{(s)}$) and their eigenvectors $\chi_n(\alpha)$ (resp. $\chi_n^{(s)}(\alpha)$) as the condensate fraction and pair wavefunctions. Moreover, we calculate the local snapshot hole densities,
\begin{align}
\label{hole_density}
n_h(\mathbf{r}) &= 1 - \langle n(\mathbf{r})\rangle_T,\\
\label{hole_density_snapshot}
n_h^{(s)}(\mathbf{r}) &= 1 - \langle\psi_s|n(\mathbf{r})|\psi_s\rangle.
\end{align}

\begin{figure*}[t!]
\vspace{-0cm}
\includegraphics[width=1.0\textwidth,clip=true]{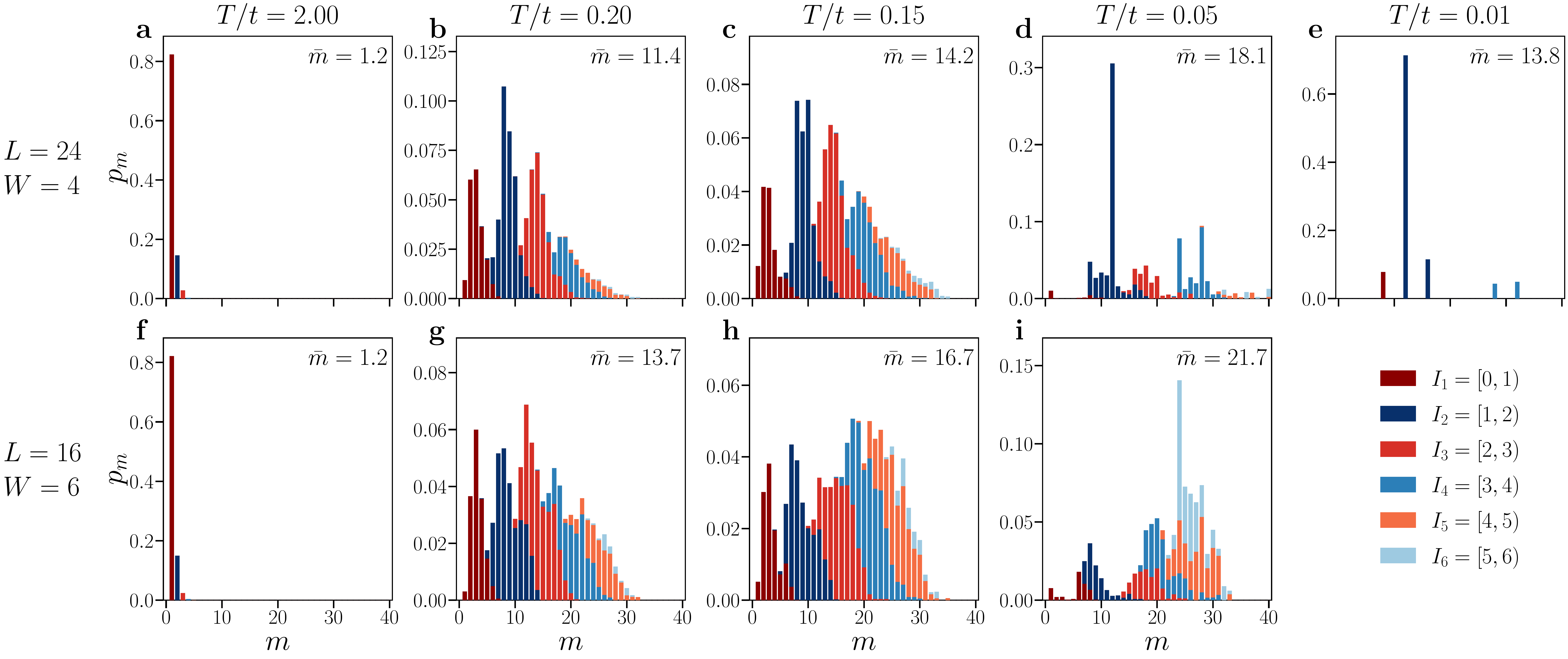}
\vspace{-0.2cm}
\caption{\textbf{Forestalled phase separation from hole-cluster statistics.}
Density-weighted cluster-size distributions $p_m$ extracted from individual METTS snapshots using the on-site hole density
$n_h^{(s)}(\mathbf r)$ [Eq.~\eqref{hole_density_snapshot}], and the cluster-identification protocol of Ref.~\cite{Sinha2025} for doping $p=1/16$, $J/t=0.4$ and $t'/t=0.2$.
Panels \textbf{(a--e)} show $24\times4$ cylinders at $T/t=2.00,\,0.20,\,0.15,\,0.05,\,0.01$, and panels \textbf{(f--i)} show $16\times6$ cylinders at $T/t=2.00,\,0.20,\,0.15,\,0.05$ (left to right).
Bars at fixed $m$ are stacked by cluster hole-mass bins $I_{k}=[k-1,k)$; the total stacked height at fixed $m$ is $p_m=\sum_{k\ge 1}p_m^{(k)}$ [Eq.~\eqref{eq:clust_pm_binned}].
The mean cluster size $\bar m$ is annotated in each panel.
Cooling transfers weight from $m=1$ to larger sizes and increases $\bar m$, consistent with forestalled phase separation. We find a stripe-like reorganization at the lowest shown temperature, $T/t=0.01$, on $24\times4$ with most weight concentrated on $m=12$.}

\label{fig:4}
\end{figure*}
\noindent \textbf{Snapshot charge densities and pairing.-} To illustrate the generic behavior, we show the properties of one representative METTS snapshot in Fig.~\ref{fig:2} for a cylinder of size $L\times W=16\times6$ at $T/t=0.04$.
Panel (a) shows the snapshot singlet 2RDM spectrum $\epsilon_n^{(s)}$ exhibiting three clearly separated eigenvalues from the rest of the spectrum indicating three relevant pair wavefunctions. Panels (b--d) display the corresponding three leading eigenvectors $\chi_n^{(s)}$ on the same METTS snapshot. There, $n_h^{(s)}(\mathbf{r})$ is indicated by the radius and colors of the gray disks. The real-space eigenvectors show that the dominant pair wavefunctions are not spatially uniform, but instead concentrate on the hole-rich regions.

A ground-state DMRG study for these parameters reported~\cite{Wietek2022FragmentedPRL} three dominant pairing eigenvalues, separated from the rest of the spectrum. At the same time, it was shown that the number of dominant eigenvalues corresponds exactly to the number of charge density wave maxima. According to Ref.~\cite{leggett2006quantum}, a condensate is called \textit{fragmented} if more than one leading eigenvalue scales with the number of particles. As such, the observation of multiple dominant eigenvalues was interpreted as the charge density wave causing a fragmentation of the condensate~\cite{Karlsson2026CooperCondensation}. See Appendix~\ref{app:snapshots} for other snapshots for different temperatures and system sizes.

\noindent \textbf{Fragmented superconductivity at finite temperature.-}
We now analyze the ensemble-averaged singlet 2RDM
$\rho_2(\alpha,\alpha')$ in Fig.~\ref{fig:3}. The onset of pairing occurs at temperatures where the leading eigenvalues begin to separate from the rest of the spectrum which is observed for $T/t\lesssim 0.2$ in panels~\ref{fig:3}(a,b). At lower temperatures, Fig.~\ref{fig:3}(b) shows that for a $24\times4$ cylinder, the leading set $\epsilon_{1,2,3}$ becomes clearly separated from the sub-leading eigenvalues shown as the gray tail   
($\epsilon_{n\ge 4}$). This indicates a finite-$T$ condensate fragmentation.
On a $16\times6$ cylinder, [Fig.~\ref{fig:3}(a)] the same tendency is visible at accessible temperatures. 

Interestingly, the split of the leading eigenvalues $w = \epsilon_1-\epsilon_3$ is tied to the strength of correlations between the stripe segments. For the case of a ground state superconductor fragmented by a charge density wave, the leading eigenvalues can be assigned a momentum in units of the charge density wave (CDW) unit cell and can therefore be regarded as a Bloch wave of Cooper pairs tunneling through the superlattice given by the CDW. In this case, we refer to $w$ as the \textit{bandwidth} and Ref.~\cite{Karlsson2026CooperCondensation} derived that it is determined by a product of two factors,
\begin{equation}
w = 2\,\bar{\epsilon}\,G,
\end{equation}
where $\bar{\epsilon}$ captures the local condensate strength per stripe and $G$ denotes the strength of the inter-stripe tunneling. As such, the bandwidth $w$ is a direct measure of the strength of pairing correlations between the stripes.

The temperature dependence of $w = \epsilon_1-\epsilon_3$ is shown in the insets of Fig.~\ref{fig:3} (a) and (b). We observe, that the $w$ remains close to zero above temperatures $T/t \gtrsim 0.2$. Coincidentally, we observe in the snapshots that the leading pairing wave functions are tightly localized on hole clusters. As temperatures are lowered below $T/t \lesssim 0.2$ we measure a gradual increase in $w$ consistent with the observation that pair wave functions increasingly develop support on multiple charge clusters in the snapshots until the pair wave functions become fully delocalized in the ground state. Upon cooling, the Cooper pairs tunnel through the half-filled, antiferromagnetic Mott barriers separating the hole-rich regions, coupling the condensates to form coherent waves delocalized across the system. 

In Fig.~\ref{fig:2}, at $T/t = 0.04$, the larger of the two hole-rich clusters shows internal structure consistent with two merged cluster peaks and supports two condensates: one uniform and one sign-changing. The local condensates of these two cluster peaks have hybridized into a uniform mode ($k_x = 0$) and a sign-changing mode ($k_x = \pi$). This suggests that intermediate temperatures host a mixture of condensates locally confined to isolated clusters and hybridized condensates on adjacent or merged clusters. Thus, $w$ is interpreted as a measure of the strength of this hybridization.
Fig.~\ref{fig:3} (c–e) shows the ensemble averaged charge densities and pair wave functions at $T/t = 0.005$. We observe three CDW maxima. $\chi_1$ displays uniform d-wave superconductivity spanning the entire system, whereas $\chi_{2,3}$ are modulated with non-zero momentum in units of the CDW unit cell along $\hat{x}$-direction.

\noindent \textbf{Charge clustering analysis.-} We quantify the charge clustering following Ref.~\cite{Sinha2025}. For any site-resolved scalar observable $f(\mathbf r)$,
bars and variances refer to uniform spatial averages within a single snapshot,
\begin{equation}
\overline{f}\equiv \frac{1}{N_{\rm site}}\sum_{\mathbf r} f(\mathbf r),
\qquad
\sigma_{f}^2\equiv \overline{f^2}-\overline{f}^{\,2}.
\label{eq:spatial_bar_def}
\end{equation}
To define hole clusters, we introduce the snapshot-dependent threshold,
\begin{equation}
n_h^{\rm th}\equiv \overline{n}_h+c\,\sigma_{n_h},
\label{eq:clust_hth}
\end{equation}
where we choose $c=0.5$, although other values of $c$ were checked for consistency of our results. Now we define the indicator,
\begin{equation}
\eta(\mathbf r)\equiv 
\begin{cases}
1 \qquad \text{for} \qquad n_h(\mathbf r) \geq n_h^{\rm th}\\
0 \qquad \text{else}.
\end{cases}
\label{eq:clust_eta}
\end{equation}
Thus, $\eta(\mathbf r)=1$ flags the set of hole-rich sites whose hole density exceeds the adaptive cutoff, while
$\eta(\mathbf r)=0$ elsewhere.
We then define the hole clusters $\mathcal C$ as the connected components of the set $\{\mathbf r:\eta(\mathbf r)=1\}$
under nearest-neighbor connectivity on the lattice (two sites are connected if they share an $\hat x$ or $\hat y$ bond). For each cluster we record its size,
\begin{equation}
m_{\mathcal C}\equiv |\mathcal C|,
\end{equation}
and its hole mass,
\begin{equation}
M_{\mathcal C}\equiv \sum_{\mathbf r\in\mathcal C} n_{h}(\mathbf r),
\label{eq:clust_mass}
\end{equation}
which measures how much doped charge resides in that connected hole-rich object. Our primary diagnostic of cluster size is the density-weighted cluster-size distribution,
\begin{equation}
p_m\equiv
\frac{1}{\sum_{\mathcal C} M_{\mathcal C}}
\sum_{\mathcal C:\,m_{\mathcal C}=m} M_{\mathcal C},
\qquad
\sum_m p_m=1,
\label{eq:clust_pm}
\end{equation}
and its mean cluster size,
\begin{equation}
\bar m\equiv \sum_m m\,p_m.
\label{eq:clust_mbar}
\end{equation}
Weighting by $M_{\mathcal C}$ ensures that $p_m$ tracks where the doped charge resides, rather than counting weakly hole-rich
and strongly hole-rich regions on equal footing.

To determine the hole mass carried by a cluster of a given size, we further resolve $p_m$ by the cluster hole mass. We bin $M_{\mathcal C}$ into unit hole-mass intervals $I_{k}=[k-1,k)$ and define the hole-mass-resolved contributions,
\begin{equation}
p_m^{(k)} \equiv
\frac{1}{M_{\rm tot}}
\sum_{\substack{\mathcal C:\, m_{\mathcal C}=m\\ \quad M_{\mathcal C}\in I_{k}}}
M_{\mathcal C},
\label{eq:clust_pm_binned}
\end{equation}
where $M_{\rm tot}\equiv \sum_{\mathcal C} M_{\mathcal C}$ denotes the total number of clusters, $p_m=\sum_{k\ge 1} p_m^{(k)}$, and $\sum_m p_m=1$. Thus, $p_m^{(k)}$ is the probability of observing a cluster of size $m$ with approximately $k$ holes.

\begin{figure}[t!]
\vspace{-0cm}
\includegraphics[width=1.01\columnwidth,clip=true]{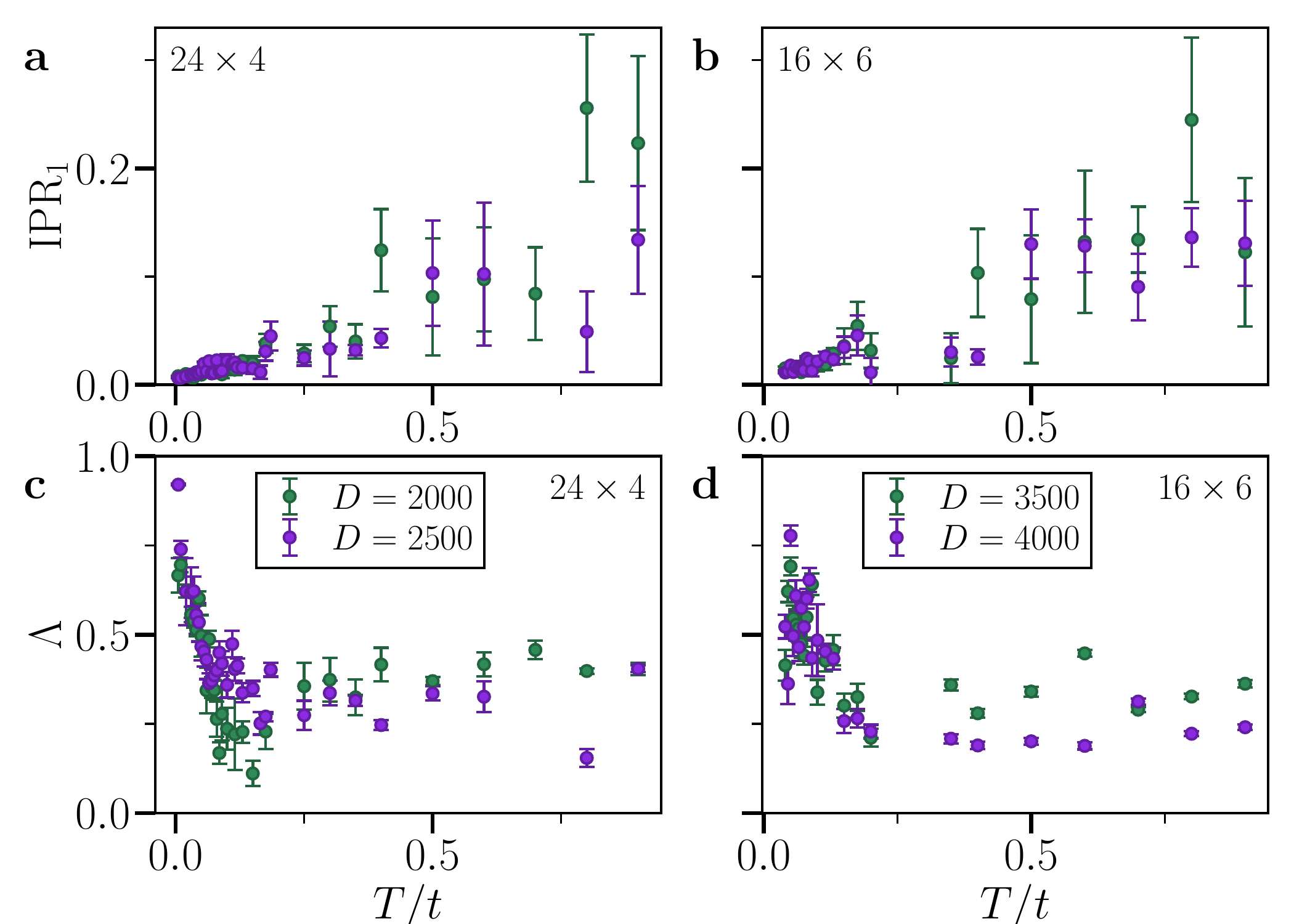}
\vspace{-0.2cm}
\caption{\textbf{Properties of the leading superconducting pair wavefunction.}
 \textbf{(a,b)} Inverse participation ratio $\mathrm{IPR}_1$ [Eq.~\eqref{eq:ipr_def}] of the leading pair wavefunction $\chi_{1}(\alpha)$ for cylinders of size $24\times4$ and $16\times6$, respectively, for doping $p=1/16$, $J/t=0.4$ and $t'/t=0.2$. \textbf{(c,d)} Pair-charge locking coefficient $\Lambda$ [Eq.~\eqref{eq:r_def}] between the on-site hole density $n_h(\mathbf r)$ and the site-centered $d$-wave amplitude $|\Delta_{d,1}(\mathbf r)|$
[Eqs.~\eqref{eq:dfield_def}--\eqref{eq:r_def}] constructed from the leading pair wavefunction. Error bars show statistical uncertainty from METTS sampling; colors compare bond dimensions.}

\label{fig:5}
\end{figure}

Figure~\ref{fig:4} summarizes the temperature evolution of the cluster-size distribution $p_m$ and its mean $\bar m$ for cylinder sizes $24\times 4$ (top row) and $16\times 6$ (bottom row). At the highest temperature, panels (a) and (f) ($T/t=2.00$) place essentially all weight at the smallest sizes ($m=1$).
Upon cooling, panels (b,c) and (g,h) ($T/t=0.20,\,0.15$) show a pronounced transfer of weight to larger $m$ and a rapid increase of $\bar m$, indicating that the doped charge increasingly resides in connected hole-rich objects extending over many lattice sites. Importantly, over this intermediate-temperature window the distributions remain broad rather than collapsing into a single large component. Intuitively, the oscillatory lobe structure in the mass-resolved stacked histograms reflects stepwise charge aggregation: as clusters grow, the dominant contribution shifts through successive hole-mass windows $I_k$. On further cooling to $T/t=0.01$ [panel~(e)], we observe the onset of stripe order: the stacked distribution is then dominated by the mass bin $I_2=[1,2)$. On $16\times6$ cylinders, panels (g--i) show the same overall shift toward larger $\bar m$ down to the lowest accessible temperature (panel (i), $T/t=0.05$), showing that hole clustering is not only particular to the $W=4$ geometry. 

\noindent \textbf{Pair-charge locking.-} Having established pairing and clustering properties individually, we now investigate their interrelation using two diagnostics. First, we investigate the localization of the leading pair wave function $\chi_1(\alpha)$, and then correlate it to the local hole density. 

\textit{(i) Localization of the dominant wavefunction.}
We measure the extent of $\chi_1(\alpha)$ in real space through its inverse participation ratio,
\begin{equation}
\mathrm{IPR}_1\equiv \sum_\alpha |\chi_1(\alpha)|^4,
\qquad
\sum_\alpha |\chi_1(\alpha)|^2=1.
\label{eq:ipr_def}
\end{equation}
For an eigenvector uniformly spread over $N_b$ bonds, one has $\mathrm{IPR}_1 = 1/N_b$, whereas an eigenvector concentrated on a small subset of bonds has a larger $\mathrm{IPR}_1$ (with $\mathrm{IPR}_1\le 1$). Figures~\ref{fig:5}(a,b) show that $\mathrm{IPR}_1$ is smallest at low temperature and increases upon heating for both
$24\times4$ and $16\times6$, indicating that the pairing correlations shorten with increasing temperature, and at very low temperatures, it approaches with $1/N_{b}$, indicating complete delocalization over the full system. 

\textit{(ii) Co-localization with hole clusters.}
To quantify whether the dominant pair wavefunction preferentially resides on hole clusters, we define the snapshot site-centered $d$-wave amplitude,
\begin{equation}
\begin{aligned}
\Delta_{d,1}^{(s)}(\mathbf r)\equiv \frac{1}{4}\Big[
&\chi_1^{(s)}(\mathbf r,\hat x)+\chi_1^{(s)}(\mathbf r,-\hat x)\\
-&\chi_1^{(s)}(\mathbf r,\hat y)-\chi_1^{(s)}(\mathbf r,-\hat y)
\Big].
\end{aligned}
\label{eq:dfield_def}
\end{equation}
We compare its magnitude with the snapshot hole-density $n_h^{(s)}(\mathbf r)$ by introducing the correlation coefficient,
\begin{equation}
\mathrm{corr}(X,Y)\equiv
\frac{\overline{XY}-\overline{X}\,\overline{Y}}
{\sqrt{\big(\overline{X^2}-\overline{X}^2\big)\big(\overline{Y^2}-\overline{Y}^2\big)}} ,
\label{eq:corr_def}
\end{equation}
 and defining the pair-charge locking coefficient,
\begin{equation}
\Lambda\equiv \mathrm{corr}\!\Big(n_h^{(s)}(\mathbf r),\,|\Delta_{d,1}^{(s)}(\mathbf r)|\Big).
\label{eq:r_def}
\end{equation}
with $\Lambda\in[-1, 1]$. Thus $\Lambda>0$ means that the envelope of the leading pair wavefunction is enhanced in hole-rich regions,
$\Lambda\simeq 0$ indicates no systematic co-localization, and $\Lambda<0$ indicates anticorrelation. Figures~\ref{fig:5}(c,d) show that $\Lambda$ is strongly positive at low temperatures and decreases upon heating, consistent
with the gradual loss of strong hole clustering.
Taken together, the simultaneous rise of $\mathrm{IPR}_1(T)$ and decrease of $\Lambda$ with $T$ provide a direct quantitative
bridge between the charge clustering trends in Fig.~\ref{fig:4} and the pairing phenomenology in Figs.~\ref{fig:2} and
\ref{fig:3}: the dominant pair wavefunction is strongly locked to hole clusters.

\begin{figure}[t!]
\vspace{-0cm}
\includegraphics[width=0.87\columnwidth,clip=true]{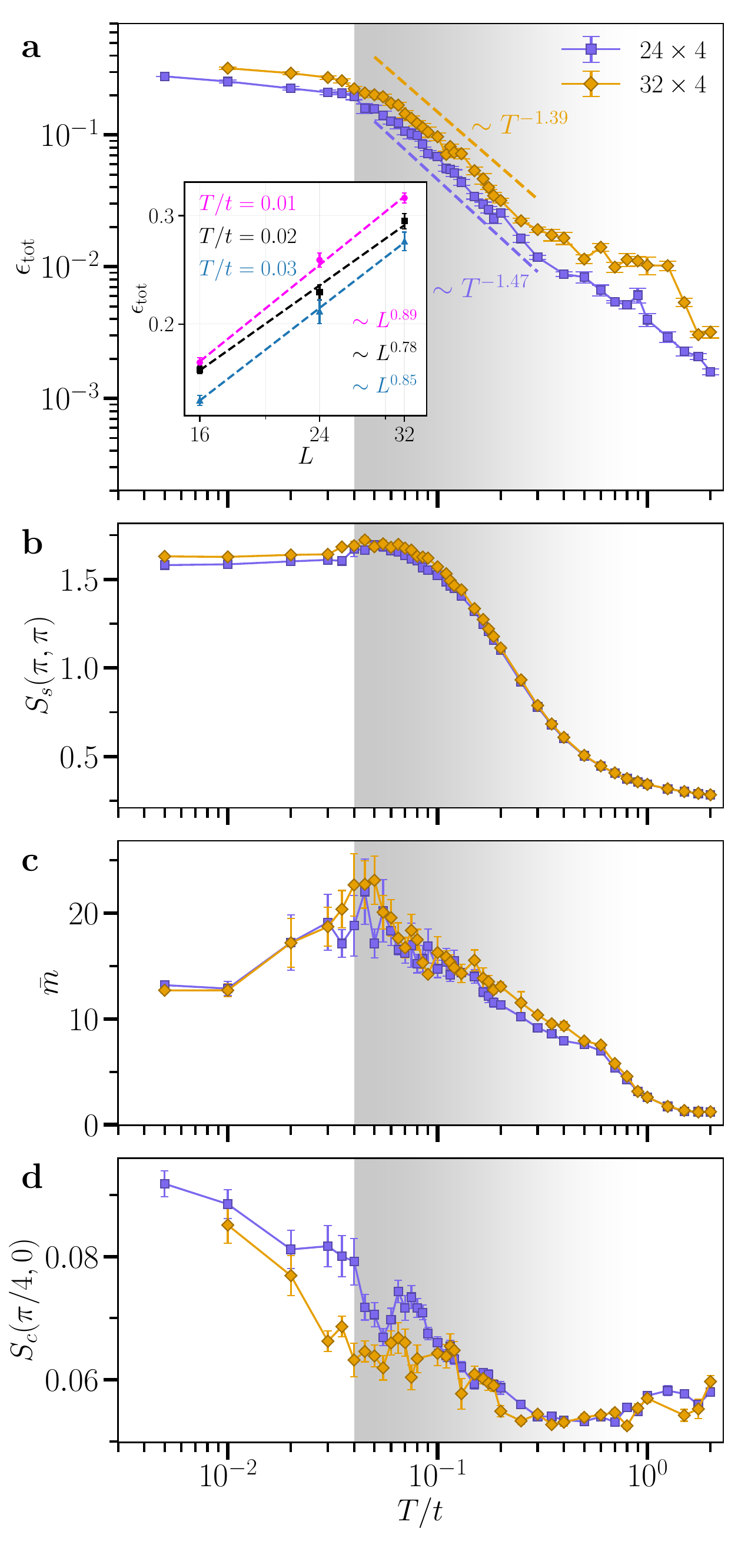}
\vspace{-0.2cm}
\caption{\textbf{Joint temperature evolution of pairing, magnetism, clustering, and stripe correlations.}
Comparing temperature dependence of four observables on $24\times4$ and $32\times4$ cylinders at doping $p=1/16$, $t'/t=0.2$, $J/t =0.4$:
(a) sum of dominant pairing eigenvalues $\epsilon_{\text{tot}}$ of the thermal pairing density matrix $\rho_2$; it grows as a power law of $(T/t)^{-\eta}$ at intermediate temperatures. The insets show algebraic scaling as a function of the cylinder length $L$, consistent with the formation of (quasi-)condensates.
(b) antiferromagnetic correlations quantified by the equal-time spin structure factor $S_s(\pi,\pi)$
[Eq.~\eqref{eq:spin_sf}];
(c) mean hole cluster size $\bar m$ [Eq.~\eqref{eq:clust_mbar}];
(d) charge structure factor at momenta  $k_x=\pi/4$ (proxy for period-$8$ stripe correlations along the cylinder axis).
The shaded region, $0.03\le T/t \le 0.15$, marks a temperature window where AFM correlations are already
substantial and charge organization is pronounced (enhanced $\bar m$ and a shoulder in $S_c(\pi/4,0)$), while
the leading pairing eigenvalue $\epsilon_{\text{tot}}$ remains strongly enhanced compared to higher temperatures.}
\label{fig:6}
\end{figure}

\noindent \textbf{Connecting pairing, antiferromagnetism, clustering, and stripes.-} We study the antiferromagnetism through the static spin structure factor,
\begin{equation}
S_s(\mathbf k)
=
\frac{1}{LW}\sum_{\mathbf r,\mathbf r'}
e^{i\mathbf k\cdot(\mathbf r-\mathbf r')}
\langle S^z(\mathbf r)S^z(\mathbf r')\rangle_T,
\label{eq:spin_sf}
\end{equation}
at wave vector $\mathbf{k}=(\pi,\pi)$. To quantify charge orders, we also consider the static charge structure factor,
\begin{equation}
S_c(\mathbf{k}) = \frac{1}{LW} \sum_{\mathbf r,\mathbf r'} e^{i\mathbf{k} \cdot (\mathbf{r} - \mathbf{r'})} \langle (n(\mathbf{r}) - n)(n(\mathbf{r'}) - n) \rangle_T
\label{eq:charge_sf}
\end{equation}
where $n$ denotes the average electron density. We focus on the longitudinal cut $S_c(k_x,0)$ and use the peak at the stripe wavevector $k_x=\pi/4$ as a proxy for
period-$8$ stripe correlations along the cylinder axis (Ref.~\cite{Wietek2022FragmentedPRL} establishes charge stripe of wavelength $8$ as the ground state order for the exact parameters). Additionally, we calculate the total condensate spectral weight,
\begin{equation}
    \epsilon_{\text{tot}}=\sum_{n=1}^{k(L)}\epsilon_n,
\end{equation}
where $k(L)$ is the number of dominant eigenvalues for a length $L$ cylinder, i.e., $k(24)=3$ and $k(32)= 4$.

These quantities are shown along with the mean cluster size $\bar{m}$ in Fig.~\ref{fig:6}. All observables start growing continuously once AFM correlations start building up at $T/t\approx 1$. We identify a characteristic temperature $T/t \approx 0.03$ where stripe order sets in, $\epsilon_{\text{tot}}$ exhibits a shoulder, and $\bar{m}$ starts decreasing after reaching a 
maximum. Below this $T$, we observe algebraic scaling $\epsilon_{\text{tot}} \propto L^\nu$ (inset of panel \ref{fig:6}(a)), with exponents $0.78 \leq \nu \leq 0.89$ consistent with the Penrose-Onsager criterion for (quasi-)condensates as expected for a quasi-one-dimensional system~\cite{Karlsson2026CooperCondensation}. We further identify a range of temperatures above $T/t = 0.03$ (shaded region), where we find that the
system is simultaneously AFM-correlated and charge-clustered: (b) the AFM structure factor $S_s(\pi,\pi)$ is growing, and (c) the mean cluster size $\bar m$ is enhanced. In this intermediate temperature range, (a) the growth of $\epsilon_{\text{tot}}$ is consistent with a power law: $\epsilon_{\text{tot}} \sim (T/t)^{-\eta}$. Interestingly, we find $\eta \approx 1.47$, comparable to the temperature scaling of the pairing susceptibility in Ref.~\cite{qu2024prl} for cylinders of width $W=4$. The charge structure factor $S_c(\pi/4,0)$ develops a visible shoulder around the same temperature as the bandwidth $w$ starts to grow (cf. Fig.~\ref{fig:3} insets). It then rises rapidly below $T/t\sim 0.03$, consistent with the strengthening of stripe order at low temperature. A complementary momentum-space view of the clustering-to-stripe crossover, and of how the leading pair wavefunctions evolve, is given in Appendix~\ref{app:kspace}.

\noindent \textbf{Discussion.-} The $t$--$t'$--$J$ model is among the most minimalistic models to describe cuprate superconductors. As such, it is natural to ask whether our result when solving the model on cylindrical geometries captures experimental observations in material compounds. 

Since our findings are based on analysis of real-space quantities, it is natural to connect with scanning tunneling experiments (STM) of cuprate superconductors. Clear experimental STM evidence for pronounced charge inhomogeneities has been found in Bi$_2$Sr$_2$CaCu$_2$O$_{8+x}$ 
(BSCCO)~\cite{Pan2001}, where it was argued that the inhomogeneities are not due to disorder but an intrinsic electronic property. Our results now suggest that already the simple $t$--$t'$--$J$ model is able to correctly model this ``nanoscale'' phase separation. Interestingly, Ref.~\cite{Pan2001} also performed a cross-
correlation analysis (cf. Fig. 2 therein) between the local density of states (LDOS) and the superconducting gap and arrive at the conclusion that the LDOS and superconducting gap are tightly spatially correlated. This experimental analysis 
is analogous to our analysis evaluating the pair-charge locking coefficient $\Lambda$ in 
Fig.~\ref{fig:5}, also demonstrating a tight correlation between charge clusters and the support of the pair wave functions. We find significant pair-charge locking is already observed at temperatures above the stripe regime. In agreement with our findings, Ref.~\cite{Gomes2007} reported the formation of nanometer-size pairing regions at temperatures above the superconducting $T_c$ in BSCCO using STM. Ref.~\cite{Kohsaka2007} also reports similar findings for the lightly doped Ca $_{1.88}$Na$_{0.12}$CuO$_2$Cl$_2$, where the observed charge inhomogeneities have been described as an intrinsic electronic glass. Further findings of charge cluster of ``puddle'' formation using STM have been reported in Refs.~\cite{tromp2023puddle, Li2021}. Thus, the phenomenon of nanoscale phase separation and the formation of local pairing above the superconducting $T_c$ is widely observed throughout the STM literature. As reported in Ref.~\cite{Sinha2025}, forestalled phase separation and charge clustering also occur in the Hubbard model, and we expect that similar phenomena are generic to a wide range of strongly correlated electron models at intermediate temperatures above the ground state regime. 

Besides STM measurements, nuclear magnetic resonance (NMR) experiments have also repeatedly discussed the possibility of nanoscale phase separation in cuprates. In the electron-doped cuprate Nd$_{1.85}$Ce$_
{0.15}$CuO$_{4-\delta}$, NMR experiments presented evidence that charge carriers form large clusters in an antiferromagnetic background~\cite{Bakharev2004}. Recent experiments using short spin-echo times on lightly-doped La$_{2-x}$Sr$_x$CuO$_4$ (LSCO) provided microscopic evidence of electronic inhomogeneity~\cite{Vuckovic2025}, which was attributed to a transition from a state with charge clusters (which the authors call ``disconnected metallic islands'') to a metallic state with tunneling between the clusters. Moreover, recent nonlinear conductivity measurements on several cuprate compounds reported superconducting precursors above $T_c$, which the authors attribute to locally paired regions in an intrinsic inhomogeneous background~\cite{Pelc2018}.

An open question in the study of cuprate superconductors is the nature of the strange metallic regime above the superconducting dome. We would like to point out that recent mesoscopic theories of the strange metal consider the charge inhomogeneity as a crucial ingredient to explain strange metallic behavior~\cite{Pelc2019,Thornton2023,Bashan2026}. As such, our results give support to the underlying assumption of charge inhomogeneities in the doped cuprates as a possible origin of strange metal behavior.

An interesting future direction is to connect our real-space observation of paired cluster formation to momentum-space diagnostics such as the electronic Green's function measured in angle-resolved photoemission spectroscopy (ARPES). Finite-temperature dynamical spectral functions are challenging targets to simulate using tensor networks, but recent works have conclusively demonstrated the feasibility of such simulations~\cite{Wang2026a,Wang2026b}. Thereby, we expect interesting connections emerging between the cluster formation observed in STM~\cite{Pan2001,Gomes2007,Kohsaka2007,tromp2023puddle, Li2021, pasupathy2008electronic}, NMR~\cite{Bakharev2004,Vuckovic2025}, and our numerics of the $t$--$t'$--$J$ model and the phenomena typically associated with the pseudogap regime, such as the widely observed momentum differentiation in ARPES studies of pseudogap cuprates.

In summary, using METTS simulations of the two-dimensional $t$--$t'$--$J$ model on cylinders of widths $W=4,6$, we have found strong evidence for a scenario where the low-temperature superconducting stripe order emerges from a strongly correlated, charge inhomogeneous background at intermediate temperatures. In this precursor regime, doped holes form fluctuating charge clusters embedded in an antiferromagnetic parent state. Pairing is tightly localized on these hole-rich regions, demonstrated by evaluating the pair-charge locking coefficient $\Lambda$. Upon cooling into the stripe regime, coherence between the clusters is increasingly established until at zero temperature the pair-wavefunctions become completely delocalized. Thus, our results establish disordered and inhomogeneous paired hole clusters as a natural precursor state to superconducting stripes, in agreement with the widely reported nanoscale phase separation in STM and NMR studies of cuprate superconductors.


\begin{acknowledgments}
We thank Rafael Soares and Johannes Hofmann for encouraging discussions. A.S. acknowledges the Alexander von Humboldt Foundation for support under the Humboldt Research Fellowship. A.W. acknowledges support by the German Research Foundation (DFG) through the Emmy Noether program (Grant No. 509755282) and the European
Research Council (ERC) under the European Union’s Horizon Europe research and innovation program (Project ID 101220368)—ERC Starting Grant MoNiKa.
\end{acknowledgments}

\bibliography{ref.bib}
\newpage

\appendix

\section{Size dependence and next-nearest neighbor coupling of the dominant pairing sector on $W{=}4$}\label{app:dom_sector}

\begin{figure*}[t]
  \centering
  \includegraphics[width=\textwidth]{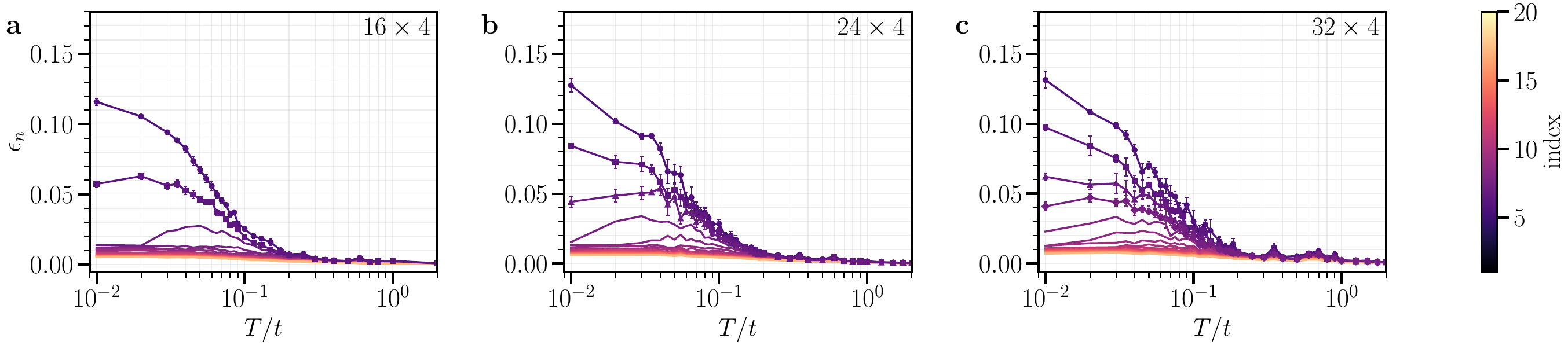}
  \caption{\textbf{Size dependence of the dominant pairing sector on $\mathbf{W=4}$}.
 In (a-c), largest $20$ eigenvalues $\epsilon_n$ of the thermally averaged pairing density matrix $\rho_2$ for cylinders of size
  $16\times4$, $24\times4$, and $32\times4$. These panels show directly that increasing $L$ in units of the charge density wavelength increases the number of leading eigenvalues
  that remain simultaneously large and well separated from the bulk at low temperature.}
  \label{fig:app:dom_sector}
\end{figure*}

In the ground state, DMRG for the $t$--$t'$--$J$ model on a $W{=}4$ cylinder at
doping $p=1/16$, $t'/t=0.2$, and $J/t=0.4$ finds stripe order with wavelength $\lambda\simeq 8$ intertwined with
$d$-wave superconductivity~\cite{Wietek2022FragmentedPRL}. In this setting, the number of leading (macroscopic) 2RDM
eigenvalues has been found to correspond to the number of charge-density-wave peaks along the cylinder, i.e., it scales as $L/\lambda \approx L/8$.
For the same parameters and geometry, our finite-temperature METTS data shows the same qualitative size dependence:
increasing $L$ in units of $\lambda$ increases the number of leading pairing eigenvalues at low temperatures, see Fig.~\ref{fig:app:dom_sector}. 

Figure~\ref{fig:app:dom_sector}(a--c) shows the leading part of this spectrum (the top $20$ eigenvalues) for system sizes
$L\times W = 16\times4,\,24\times4,\,32\times4$.
The key visual message is that the count of clearly separated leading eigenvalues grows with $L$.

\begin{figure*}[t]
  \centering
  \includegraphics[width=\textwidth]{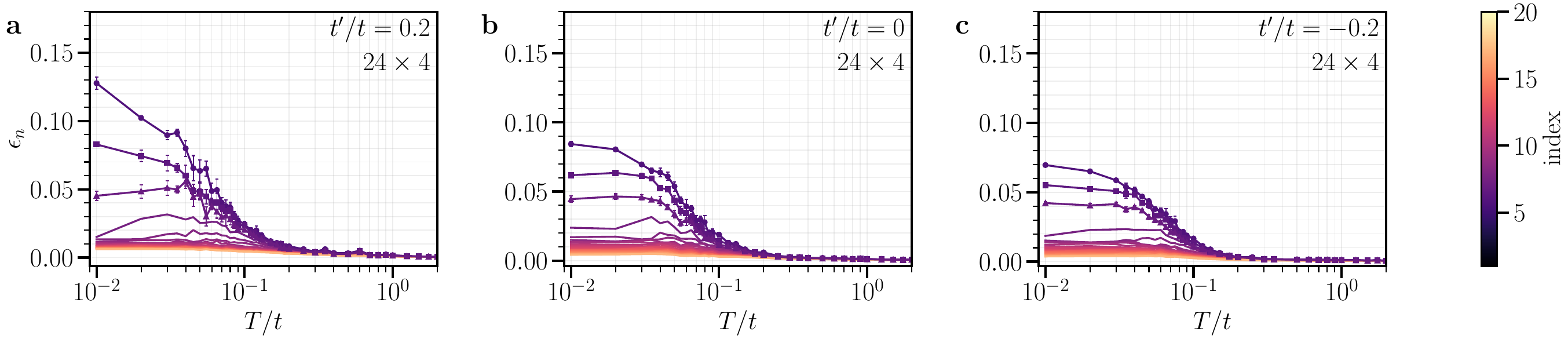}
  \caption{\textbf{Dependence of the dominant pairing sector on $\mathbf{W{=}4}$ for next-nearest-neighbor coupling $\mathbf{t'/t}$.}
  In (a--c), largest $20$ eigenvalues $\epsilon_n$ of the thermally averaged pairing density matrix $\rho_2$ for cylinder of size
  $24\times4$ for next-nearest-neighbor couplings $t'/t=-0.2, 0.0$, and $0.2$.}
  \label{fig:app:dom_tp}
\end{figure*}

Figure~\ref{fig:app:dom_tp} compares the same quantities for three values of the next-nearest-neighbor hopping, $t'/t=-0.2,\,0,\,0.2$. In all three cases, the leading positive eigenvalues grow upon cooling, showing that pairing correlations are enhanced at low temperature irrespective of the sign of $t'$. At the same time, the detailed structure of the dominant pairing sector depends noticeably on $t'$. For (a) $t'/t=0.2$, the leading eigenvalues reach the largest magnitudes and are most clearly separated from the subleading spectrum. For (b) $t'/t=0$, the same tendency remains visible but is reduced in magnitude, with a more modest low-temperature enhancement and less pronounced separation. For (c) $t'/t=-0.2$, the low-temperature growth of the leading modes is weakest and the spectrum appears more compressed overall. 

\section{Technical Details of METTS simulations}
\label{app_technical}

We study the hole-doped $t$--$t'$--$J$ model on square-lattice cylinders with open boundaries along $\hat{x}$ (length $L$) and periodic boundaries along $\hat{y}$ (circumference $W$). The Hamiltonian is given in Eq.~\eqref{eq:ttJ_ham}. Throughout the main text we set $t=1$ and focus on $J/t=0.4$ and $t'/t=0.2$ (positive $t'$, following the common electron-doped sign convention) at fixed hole doping $p=1/16$ (filling $n=1-p$). We simulate cylinders up to sizes $32\times 4$ and $16\times 6$.

Thermal expectation values are obtained using minimally entangled typical thermal states (METTS)~\cite{white2009,
stoudenmire2010minimally}. In brief, METTS generates a Markov chain of pure states whose sample average approximates the
thermal trace; the explicit definitions used in this work are given in Eqs.~\eqref{eq:metts_state} and
\eqref{eq:metts_average} in the main text. Our implementation is built on top of the public repository \texttt{ITensors.jl}~\cite{10.21468/SciPostPhysCodeb.4-r0.3,10.21468/SciPostPhysCodeb.4} and a custom METTS repo publicly available at \url{https://github.com/awietek/METTS.jl}. We work in the $t$-$J$ site basis with conserved quantum numbers during imaginary-time evolution. After each measurement we collapse the METTS to a product state by sampling in the $X$ basis. This is done to reduce autocorrelations. 

Imaginary-time evolution $e^{-\beta H/2}$ is performed using TDVP evolution of MPS. Concretely, we use a TDVP scheme with an initial subspace (basis) expansion step, implemented via a Krylov-based enlargement of the local bond space before short $1$-site TDVP updates. This preconditioning reduces sensitivity to the initial product state at larger $\beta$ and improves robustness when the relevant manifold cannot be reached by purely $1$-site updates. After this initial stage, the remaining evolution uses adaptive switching between $2$-site TDVP (to allow bond growth) and $1$-site TDVP once the maximum bond dimension is reached. The numerical controls are the imaginary-time step size $\tau$, the truncation cutoff, and the maximum bond dimension $D_{\max}$. We normalize the state during evolution. Representative convergence checks were carried out by varying $D_{\max}$ for key observables shown in the main figures. When starting the METTS chain, we optionally perform a short DMRG preconditioning run at the target
filling to generate a low-entanglement physical initial state. We used $D_{\max} = 2500$ for cylinders of width $W=4$ and $4000$ for $W=6$.

 For each temperature, we discard the first few steps as a warm-up. All observables are computed from the remaining snapshots. Uncertainties are estimated by bootstrap resampling over METTS snapshots (with replacement). To learn about more implementation details, see Ref.~\cite{wieteksquare2021}.

\section{METTS Snapshots for different parameters of the $t$--$t'$--$J$ model}
\label{app:snapshots}

Here we provide additional real-space METTS snapshots that complement the thermally averaged analysis in the main text. For each selected METTS configuration $|\psi_s\rangle$, we construct the corresponding snapshot singlet-pair density matrix $\rho_2^{(s)}$, diagonalize it, and display its eigenvalue spectrum together with the three leading pair wavefunctions $\chi^{(s)}_1$, $\chi^{(s)}_2$, and $\chi^{(s)}_3$ overlaid on the same snapshot hole-density profile $n_h^{(s)}(\mathbf r)$. The purpose of these figures is to look underneath the thermal curtain and visualize directly where the dominant pairing weight resides in a typical thermal state.

\begin{figure*}[t]
\centering
\includegraphics[width=0.89\textwidth]{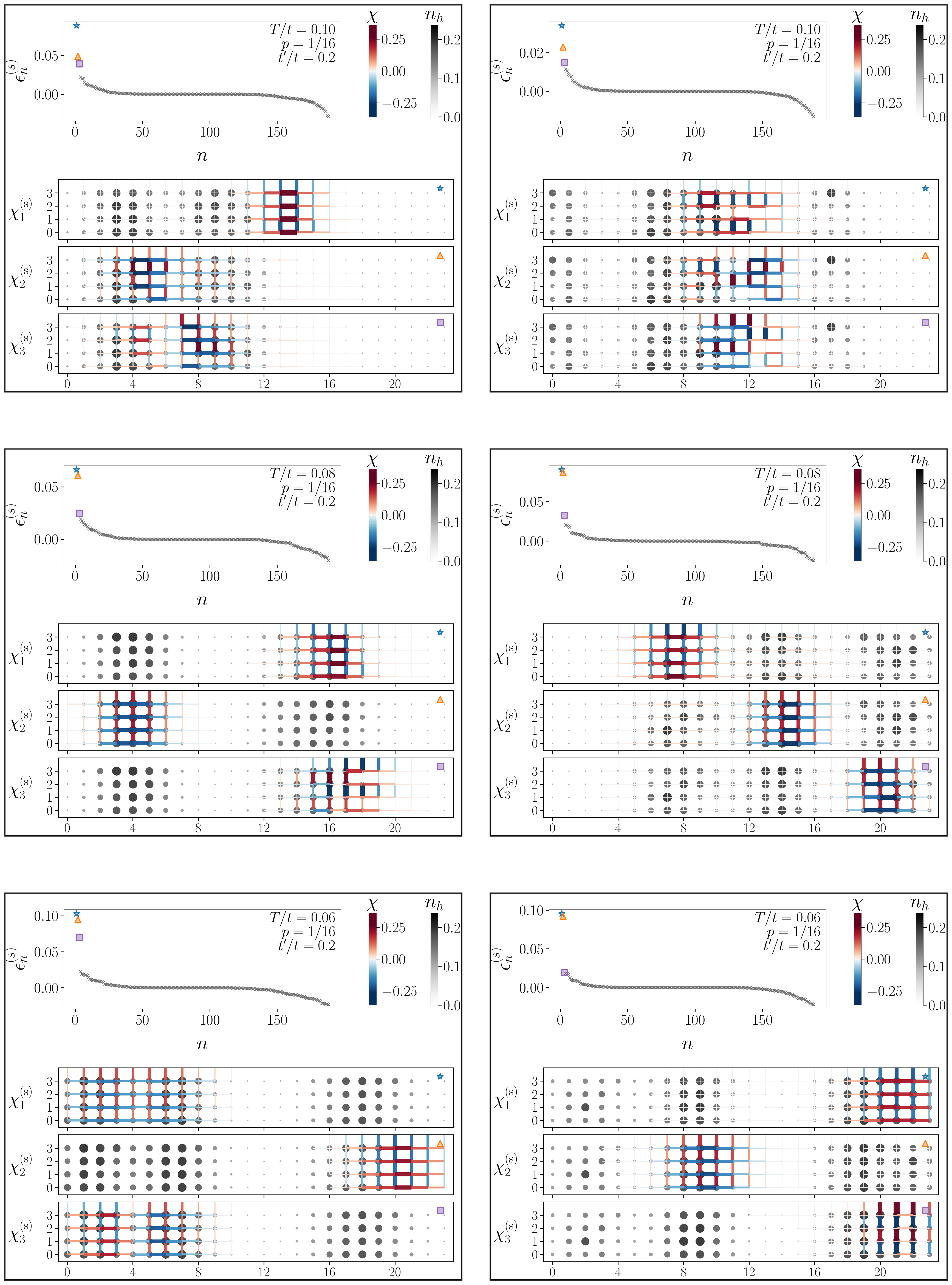}
\caption{Representative METTS snapshots on the $24\times4$ cylinder at doping $p=1/16$ and $t'/t=0.2$ for $T/t=0.10$, $0.08$, and $0.06$ (two independent snapshots for each temperature). In each block, the top panel shows the eigenvalue spectrum $\epsilon_n$ of the snapshot singlet-pair density matrix $\rho_2^{(s)}$, while the three panels below show the corresponding leading pair wavefunctions $\chi^{(s)}_{1,2,3}$ overlaid on the same snapshot hole-density profile $n_h^{(s)}(\mathbf r)$. gray circles encode the local hole density, with larger and darker circles indicating larger $n_h$, and colored bonds encode the signed bond amplitude of the pairing eigenmode. The detailed arrangement of the clusters varies from one METTS configuration to another, but the pair-charge locking is robust.
}

\label{fig:app_snapshot1}
\end{figure*}

The main message is that the dominant pairing modes are strongly tied to hole-rich regions over a broad temperature window. At intermediate temperatures on the $24\times4$ cylinder for $t'/t=0.2$ (Fig.~\ref{fig:app_snapshot1}), the leading pair wavefunctions are localized on individual hole-rich clusters or on a small set of nearby clusters, while the hole-poor antiferromagnetic background carries comparatively little pairing weight. Independent METTS snapshots at the same temperature fluctuate from sample to sample, but the geometric locking between pairing and hole-rich regions is robust.

\begin{figure*}[t]
\centering
\includegraphics[width=0.9\textwidth]{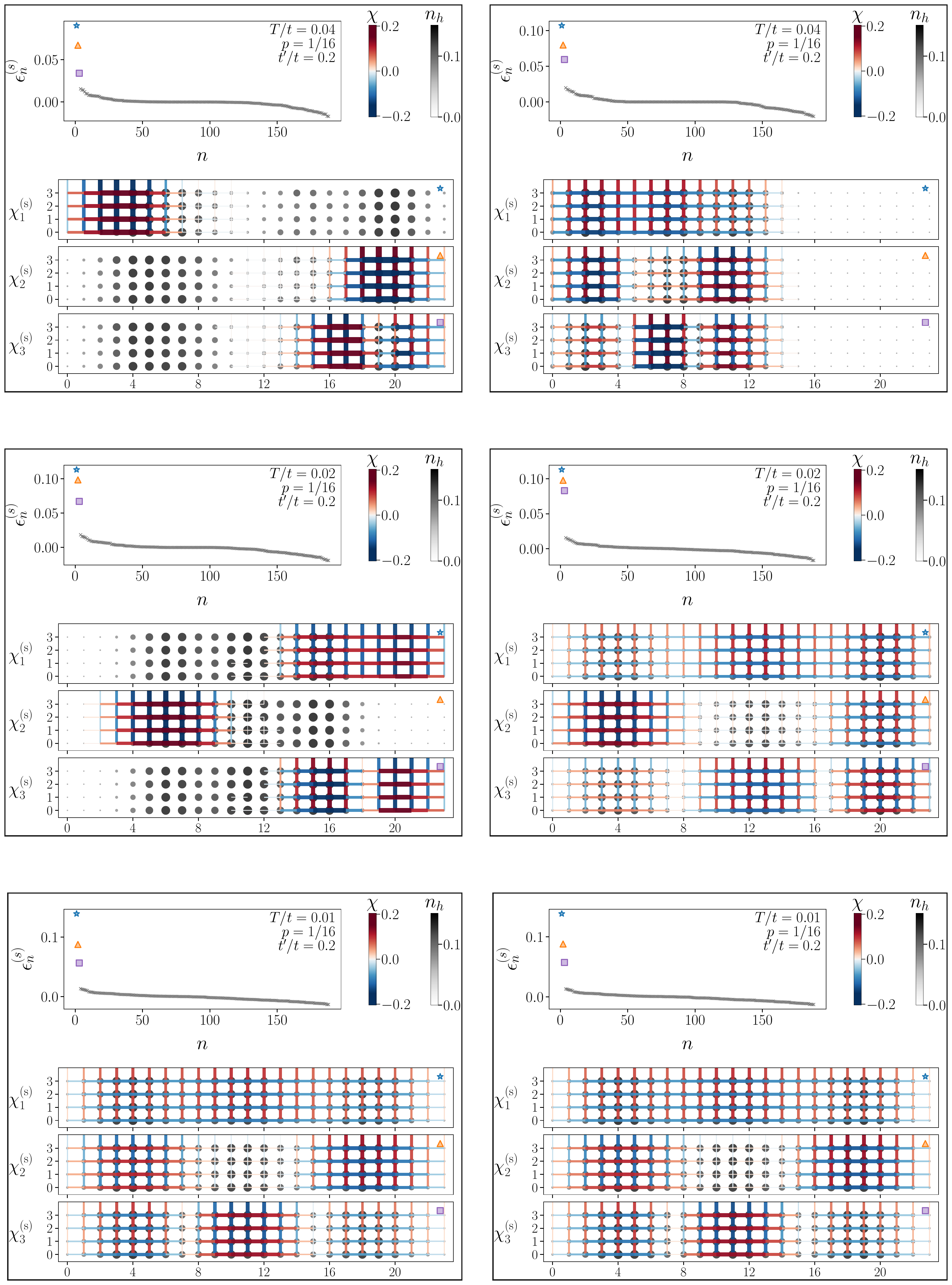 }
\caption{Representative low-temperature METTS snapshots on the $24\times4$ cylinder at doping $p=1/16$ and $t'/t=0.2$ for $T/t=0.04$, $0.02$, and $0.01$ (two independent snapshots for each temperature). Compared to the intermediate-temperature snapshots of the previous figure, the leading pairing modes become progressively more extended and strongly connected across different hole-rich regions upon cooling. At the lowest temperature, the leading pair wavefunction $\chi_1$ develops a coherent, system-spanning $d$-wave-like pattern.}

\label{fig:app_snapshot2}
\end{figure*}

\begin{figure*}[t]
\centering
\includegraphics[width=0.8\textwidth]{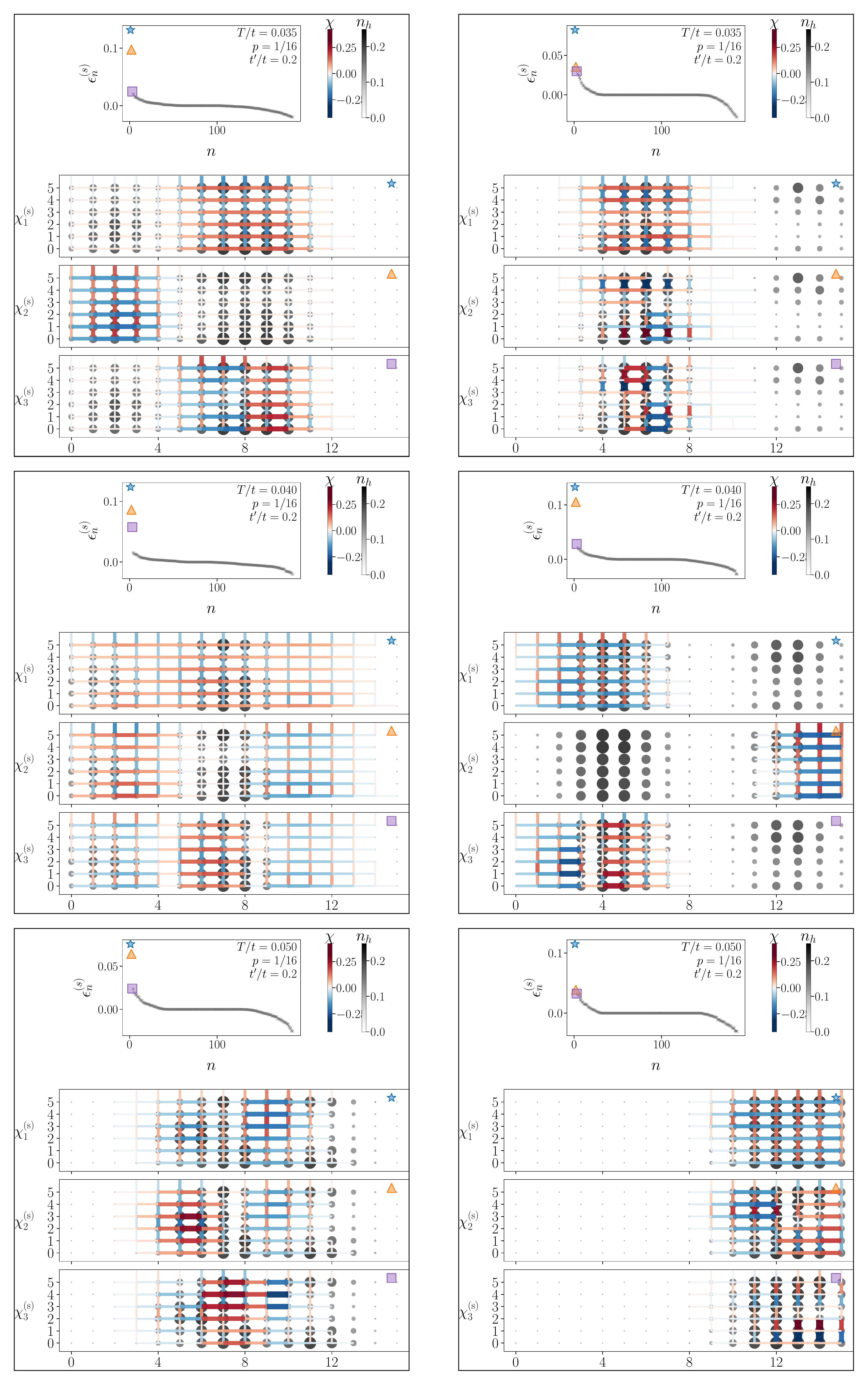}
\caption{
Representative METTS snapshots on the $16\times6$ cylinder at doping $p=1/16$ and $t'/t=0.2$ for $T/t=0.035$, $0.04$, and $0.05$ (two independent snapshots for each temperature). The plotting convention is the same as in the previous figures. The dominant pair wavefunctions again track the hole-rich regions of each snapshot, showing that pair-charge locking persists also on the wider cylinder.}
\label{fig:app_snapshot3}
\end{figure*}

\begin{figure*}[t]
\centering
\includegraphics[width=0.8\textwidth]{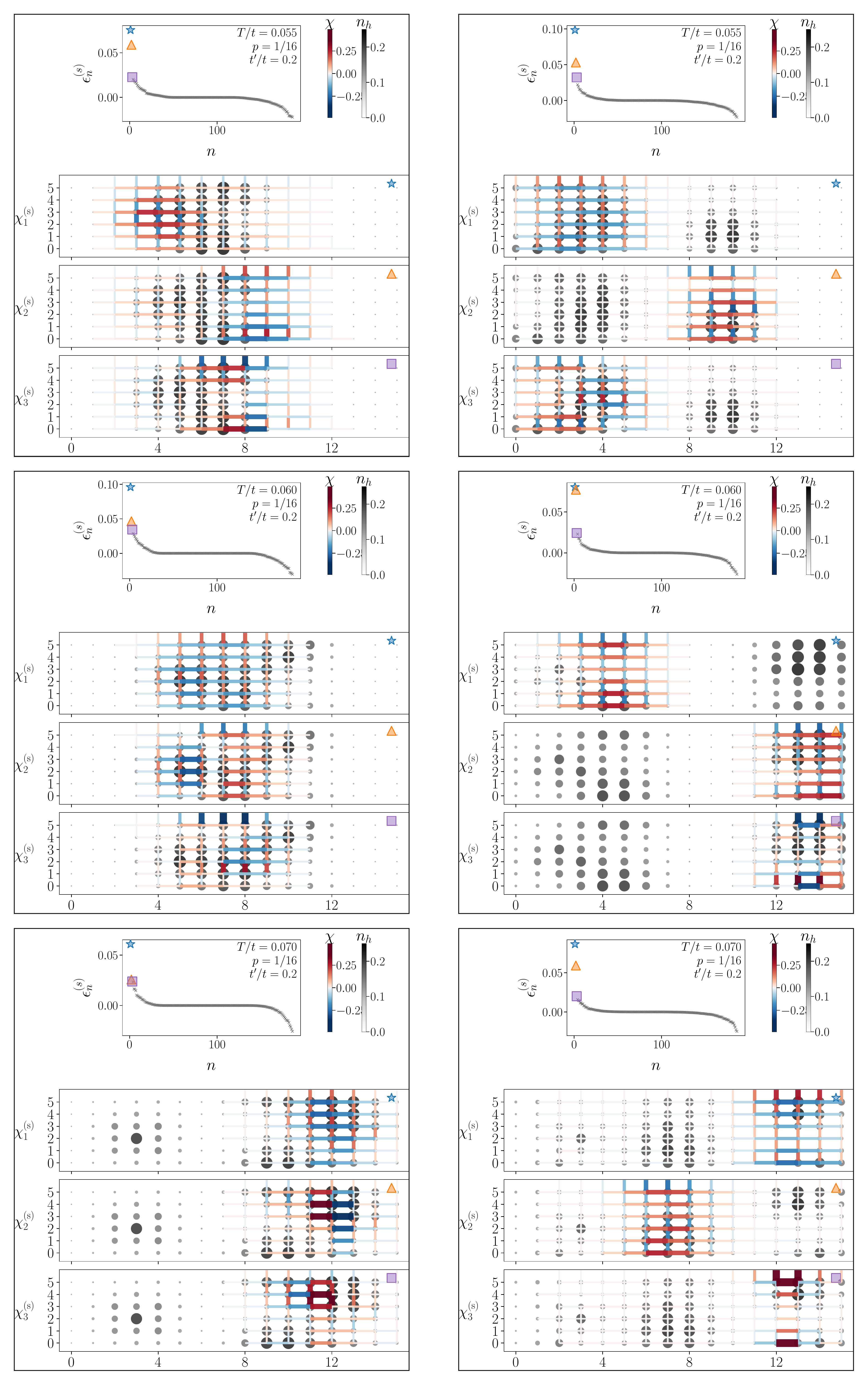}
\caption{
Representative METTS snapshots on the $16\times6$ cylinder at doping $p=1/16$ and $t'/t=0.2$ for $T/t=0.055$, $0.060$, and $0.070$ (two independent snapshots for each temperature). The plotting convention is the same as in the previous figures.}
\label{fig:app_snapshot4}
\end{figure*}

Upon further cooling (Fig.~\ref{fig:app_snapshot2}), the real-space structure of the condensates changes qualitatively. The cluster-localized pairing modes increasingly hybridize across different hole-rich regions and reorganize into stripe-compatible patterns. At the lowest temperatures shown for $t'/t=0.2$, the leading pair wavefunction $\chi^{(s)}_1$ develops a coherent, system-spanning $d$-wave-like structure, whereas the subleading wavefunctions are modulated with non-zero momentum. The corresponding $16\times6$ snapshots show the same tendency, indicating that this behaviour is not restricted to the narrower $W=4$ geometry (Figs.~\ref{fig:app_snapshot3}, \ref{fig:app_snapshot4}).

\section{Momentum-resolved charge and pairing correlations}
\label{app:kspace}

The main text diagnoses charge organization and pairing primarily in real space, using snapshot density profiles and the
pair wavefunctions of the singlet pair density matrix. Here we provide a complementary momentum-resolved view along the cylinder
axis, which is consistent with the same intermediate-temperature clustering regime and the low-temperature charge stripe regime.

\subsection*{Charge structure factor along the cylinder axis}

 In Fig.~\ref{fig:app_kspace}(a) we plot the longitudinal
cut of the charge structure factor [Eq.~\eqref{eq:charge_sf}] $S_c(k_x)\equiv S_c(\mathbf{k})$ at $\mathbf{k}=(k_x,0)$.
Because we work at fixed particle number, the strict $k_x=0$ component is not informative; on a finite cylinder the closest
proxy for a $k\to 0$ enhancement is the smallest nonzero momentum $k_x=\pm 2\pi/L$. A peak at this smallest $k_x$ therefore
indicates that the dominant charge fluctuations live on the longest available length scale, consistent with
phase-separation-like tendencies that have not reorganized into charge stripe order. Upon further cooling, the weight in $S_c(k_x)$
moves towards and sharpens near the stripe wavevector $k_x=\pm \pi/4$ (for $p=1/16$ on $L=24$), consistent with the formation
of stripe-scale charge modulations.

\subsection*{Mode-resolved $d$-wave momentum profile}

\begin{figure}[t]
\centering
\includegraphics[width=\columnwidth]{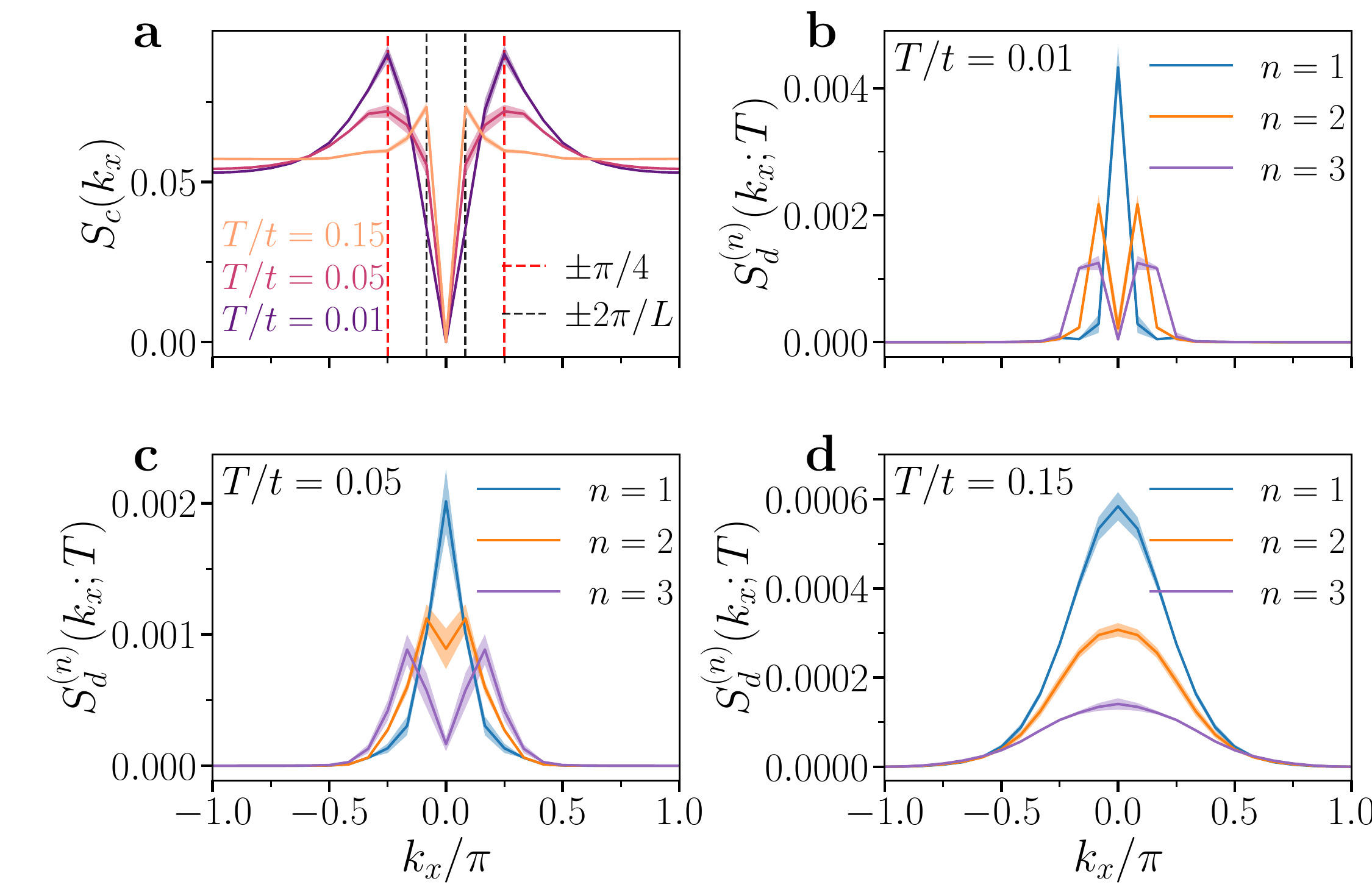}
\caption{\textbf{Momentum-resolved charge and pairing correlations.}
For the $t$--$t'$--$J$ model on a $24\times4$ cylinder at $J/t=0.4$, $t'/t=0.2$, and doping $p=1/16$, we compute: \textbf{(a)} charge structure factor $S_c(k_x)$ along the cylinder axis for $T/t=0.15,\,0.05,\,0.01$.
Vertical dashed guides mark the charge stripe wavevector $k_x=\pm \pi/4$ and the smallest nonzero momentum on the finite-$L$ grid,
$k_x=\pm 2\pi/L$.
\textbf{(b-d)} Mode-resolved $d$-wave momentum profile $S_d^{(n)}(k_x)$ for the first three pair wavefunctions
($n=1,2,3$) of the thermally averaged singlet pairing density matrix at (b) $T/t=0.01$, (c) $T/t=0.05$, and (d) $T/t=0.15$.
Shaded bands indicate statistical bootstrapping uncertainties from the underlying METTS sampling.}
\label{fig:app_kspace}
\end{figure}

To connect the pair wavefunctions to momentum space, we take the leading eigenvectors
$\chi_n(\mathbf r,\mu)$ of the thermally averaged singlet pair density matrix $\rho_2$ and map each bond field to a
site-centered $d$-wave amplitude using the same definition as in the main text~Eq.~\eqref{eq:dfield_def}. We then average over the
periodic direction,
\begin{equation}
\overline{\Delta}_{d,n}(x)\equiv \frac{1}{W}\sum_{y=1}^{W}\Delta_{d,n}(x,y),
\end{equation}
and Fourier transform along the cylinder axis,
\begin{equation}
S_d^{(n)}(k_x)\equiv
\frac{1}{L^2}\left|\sum_{x=1}^{L}\overline{\Delta}_{d,n}(x)\,e^{ik_x x}\right|^2.
\label{eq:Sd_n_def_app}
\end{equation}
Figures~\ref{fig:app_kspace}(b-d) show $S_d^{(n)}(k_x)$ for the first three pair wavefunctions at representative
temperatures. At low temperature the leading pair wavefunction is strongly concentrated near $k_x=0$, consistent with a coherent,
system-spanning $d$-wave-like structure. At higher temperature the spectra broaden and the leading modes become more
comparable in magnitude, consistent with shorter-ranged pairing.

\end{document}